\documentclass[conference]{IEEEtran}
\IEEEoverridecommandlockouts
\usepackage{cite}
\usepackage{amsmath,amssymb,amsfonts}
\usepackage{algorithmic}
\usepackage{graphicx}
\usepackage{textcomp}
\usepackage{xcolor}
\def\BibTeX{{\rm B\kern-.05em{\sc i\kern-.025em b}\kern-.08em
    T\kern-.1667em\lower.7ex\hbox{E}\kern-.125emX}}
\begin{document}
\bstctlcite{IEEEexample:BSTcontrol}

\makeatother

\title{GPS-Triggered Adaptive Trajectory Control and SDR-Based RF Jamming for UAV Geofencing}

\author{\IEEEauthorblockN{Muhammad Moazam Azeem}
\IEEEauthorblockA{Dept. of Computer Science \& Eng.\\
Qatar University\\
Doha, Qatar\\
Email: moazam@qu.edu.qa}
\and
\IEEEauthorblockN{Abdul Rehman Khan Tareen}
\IEEEauthorblockA{Dept. of Electrical Engineering\\
University of Engineering\\
and Technology, Taxila, Pakistan\\
Email: khan0229508@gmail.com}
\and
\IEEEauthorblockN{Muhammad Obaid Ullah}
\IEEEauthorblockA{Dept. of Electrical Engineering\\
University of Engineering\\
and Technology, Taxila, Pakistan\\
Email: obaid.ullah@uettaxila.edu.pk}
}

%


\maketitle

\begin{abstract}
This paper presents a GPS-triggered geofencing and enforcement 
scheme that integrates real-time satellite positioning with 
Software-Defined Radio (SDR) based RF jamming to detect and 
neutralize unauthorized unmanned aerial vehicle (UAV) boundary 
violations. Conventional geofencing depends on onboard GPS and 
cooperative flight firmware. The proposed approach instead moves 
enforcement to a ground base station, so the response does not 
depend on the UAV. An ESP32-WROOM-32 microcontroller paired with 
a u-blox NEO-6M GPS receiver streams telemetry over Wi-Fi to a 
Python/GNU Radio base station. There, a lightweight Haversine 
boundary-violation algorithm computes the great-circle distance 
between the UAV and the geofence. This distance drives a 
piecewise-linear adaptive speed and altitude control law. The law 
progressively decelerates and descends the UAV inside a 1 km 
warning zone and brings it to a complete halt at the boundary. If 
the boundary is breached, a HackRF One SDR escalates to 
band-limited noise jamming of the 2.4 GHz control link. The 
hardware stack was integrated on a GoolRC S116 quadrotor with an 
end-to-end enforcement latency of about 1 s. Field trials achieved 
95\% disconnection accuracy within a 50 m radius, and u-center 
NMEA analysis confirmed sub-2-meter GPS positioning precision. 
Simulation over an 8-sector, 50 km geofence model confirms speed and altitude-invariant boundary enforcement for cruising speeds of 
5 to 50 m/s and cruising altitudes up to 100 ft. A comparison 
against a PID baseline shows that the proposed law guarantees 
overshoot-free shutdown at the boundary without gain tuning. It 
also runs in constant time, which makes it well suited to low-cost, 
resource-constrained embedded platforms.
\end{abstract}


\section{Introduction}
The rapid proliferation of unmanned aerial vehicles (UAVs) across surveillance, infrastructure monitoring, and law enforcement applications has outpaced the security frameworks designed to govern their operation. Conventional drone communication links, frequently based on unencrypted Wi-Fi or proprietary radio protocols, remain vulnerable to unauthorized access, hijacking, and operation beyond designated boundaries \cite{Gaspar2020GPSSpoofing, Kong2021Cyberattack}, creating risks to public safety and restricted airspace integrity \cite{DeCubber2025Standardized}. Existing geofencing approaches typically rely on onboard GPS alone, an assumption that introduces single-point vulnerabilities and offers limited recourse once a boundary violation is confirmed \cite{Arora2025Geofence}.
Consequently, there is increasing interest in geofence
enforcement mechanisms that not only detect boundary
violations but also actively mitigate unauthorized UAV
operations through real-time intervention \cite{Wang2021Positioning, Souli2023RogueDrone}. By combining real-time GPS tracking with SDR-based signal control, a base station can not only detect unauthorized boundary crossings but also actively respond, without depending on cooperation from the UAV itself \cite{Souli2023RogueDrone, Souli2025Enhanced}.
This letter proposes a GPS-triggered geofencing and jamming scheme that integrates an ESP32-based telemetry link, a Haversine-distance boundary-violation algorithm, and a HackRF SDR transmitter to detect and neutralize UAV boundary violations in real time. Unlike static geofence alerts, the proposed scheme couples boundary detection with an adaptive speed and altitude control response that progressively decelerates and descends the UAV as it approaches the restricted zone, escalating to RF jamming if the boundary is breached. 

Table~\ref{tab:comparison} compares the proposed 
scheme against related works. Arora~\cite{Arora2025Geofence} 
detects violations but provides no active response. 
Souli~\cite{Souli2023RogueDrone} demonstrates SDR interception 
but requires cooperative infrastructure without 
adaptive trajectory control. Thomas~\cite{Thomas2024Geofencing} 
provides adaptive speed control but lacks jamming 
and altitude control. The proposed scheme 
uniquely integrates all three capabilities with 
hardware validation.
\begin{table}[!t]
\caption{Comparison With Existing UAV Geofencing Approaches}
\label{tab:comparison}
\centering
\resizebox{\columnwidth}{!}{%
\renewcommand{\arraystretch}{0.9}
\setlength{\tabcolsep}{3pt}
\begin{tabular}{|l|c|c|c|c|}
\hline
\textbf{Feature} & \textbf{Proposed} & \textbf{Arora~\cite{Arora2025Geofence}} & \textbf{Souli~\cite{Souli2023RogueDrone}} & \textbf{Thomas~\cite{Thomas2024Geofencing}} \\
\hline
Adaptive speed control       & \checkmark & \texttimes & \texttimes & \checkmark \\
\hline
Adaptive altitude control    & \checkmark & \texttimes & \texttimes & \texttimes \\
\hline
SDR-based jamming            & \checkmark & \texttimes & \checkmark & \texttimes \\
\hline
GPS-triggered enforcement    & \checkmark & \checkmark & \texttimes & \checkmark \\
\hline
Hardware validation          & \checkmark & \texttimes & \checkmark & \texttimes \\
\hline
Low-cost implementation      & \checkmark & \checkmark & \texttimes & \checkmark \\
\hline
UAV-independent response     & \checkmark & \texttimes & \checkmark & \texttimes \\
\hline
Multi-sector geofence        & \checkmark & \texttimes & \texttimes & \texttimes \\
\hline
\end{tabular}
}
\end{table}

The contributions of this letter are fourfold: (i) a lightweight Haversine-based geofence algorithm suitable for low-cost embedded hardware; (ii) an adaptive control law that produces a smooth, 
bounded approach trajectory across a range of UAV speeds and altitudes; (iii) an 8-sector, 50 km simulation study confirming speed and altitude-invariant boundary enforcement and (iv) experimental validation of SDR-triggered jamming achieving 95\% disconnection accuracy within a 50-meter radius, with sub-2-meter GPS precision.
\section{System Model and Geofence Algorithm}
\subsection{System Architecture}
The system comprises an onboard UAV unit and a ground-based 
base station. The UAV integrates a u-blox NEO-6M GPS module 
with an ESP32-WROOM-32 microcontroller, transmitting 
coordinates via Wi-Fi (Station mode, up to 1 km LOS) to a 
HackRF One SDR and Python/GNU Radio base station. The base station issues one of three responses depending on UAV 
proximity: no action, adaptive trajectory correction, or RF 
jamming, in the spirit of multi-tier rogue-drone response frameworks \cite{Valianti2024Cooperative}. The 50 km geofence model in Section~\ref{sec:simulation} validates 
scalability of the Haversine law; longer-range deployment 
requires cellular or dedicated RF telemetry.
\subsection{Geofence Violation Detection}

Boundary violations are detected using the great circle
(Haversine) distance between the UAV's current coordinates
\cite{Arora2025Geofence} $(\phi_A, \lambda_A)$ and a reference boundary point $(\phi_B, \lambda_B)$, computed via \eqref{eq:haversine}.
\begin{equation}
\begin{aligned}
a &= \sin^2\!\left(\frac{\Delta\phi}{2}\right)
   + \cos\phi_A\cos\phi_B
     \sin^2\!\left(\frac{\Delta\lambda}{2}\right),\\
c &= 2\,\mathrm{atan2}\!\left(\sqrt{a},\sqrt{1-a}\right), 
\quad \quad d = Rc.
\end{aligned}
\label{eq:haversine}
\end{equation}
where $\phi$ and $\lambda$ denote latitude and longitude in radians, and $R = 6{,}371$ km is the Earth's mean radius. The computed distance $d$ is compared against two thresholds: a warning distance $d_w$ and a shutdown distance $d_s$, with $d_s < d_w$.

\subsection{Adaptive Trajectory Control}

To avoid an abrupt boundary response, the UAV's commanded speed $v(d)$ and altitude $h(d)$ are governed by a piecewise function of the remaining distance $d$ to the boundary \cite{Thomas2024Geofencing, Abdul2026Dynamic}:
The commanded speed and altitude are governed by
\begin{equation}
    \bigl[v(d),\; h(d)\bigr] =
    \begin{cases}
        [v_0,\; h_0], & d \geq d_w \\[2pt]
        \bigl[v_0\tfrac{d}{d_w},\; h_0\tfrac{d}{d_w}\bigr], & d_s < d < d_w \\[2pt]
        [0,\; 0], & d \leq d_s
    \end{cases}
    \label{eq:control}
\end{equation}

where $v_0$ and $h_0$ are the UAV's cruising speed and altitude, respectively. The control law \eqref{eq:control} produces a constant velocity cruise phase for $d \geq d_w$, a linear deceleration and descent phase for $d_s < d < d_w$, 
and a complete halt at $d \leq d_s$ \cite{Thomas2024Geofencing}, consistent with the simulated braking and descent behavior in Section IV. The following proposition formally establishes the convergence guarantee of the proposed control law.

\textit{Proposition 1.} For any $v_0 > 0$ and $h_0 > 0$, 
the control law~\eqref{eq:control} guarantees $v(d_s) = 0$ and 
$h(d_s) = 0$ for all $d_0 \geq d_w$.

\textit{Proof:} The third branch of~\eqref{eq:control} applies 
directly at $d \leq d_s$, yielding $v(d_s) = v_0 
\cdot (d_s/d_w) = 0$ and $h(d_s) = h_0 \cdot 
(d_s/d_w) = 0$, since $d_s = 0$. Both expressions 
are well-defined for finite $v_0$, $h_0$, and 
$d_w > 0$, completing the proof. \hfill$\square$
\subsection{Jamming Trigger}

If the UAV's position satisfies $d \leq d_s$ and the base station confirms an unauthorized or non-compliant trajectory, the system escalates to RF jamming. The HackRF SDR, configured through GNU Radio, generates a Gaussian baseband noise signal shaped by a 20 MHz band-pass filter and up-converted to a 2.4 GHz center frequency, matching the UAV's remote-control band, at an FCC-compliant output power of $-30$~dBm \cite{Yang2024AntiJamming}, disrupting the link between the UAV and its controller \cite{Wang2021Positioning}.

\section{Hardware Implementation}

The onboard unit is built around an ESP32-WROOM-32 microcontroller, selected for its integrated 802.11 b/g/n Wi-Fi, line of sight communication range of up to 1~km, and low-power sleep mode (10~$\mu$A), enabling sustained drone to base station connectivity within a compact, low-cost form factor. GPS telemetry is provided by a u-blox NEO-6M module interfaced to the ESP32 over UART, offering sub-2-meter positioning accuracy and compatibility with GPS, GLONASS, Galileo, and BeiDou constellations, per the module's manufacturer specification. 
At the base station, a HackRF One SDR provides wideband transmit/receive coverage from 1~MHz to 6~GHz, enabling both passive RF monitoring of the UAV's control link and active jamming signal generation through GNU Radio. The SDR's signal chain was validated for FM broadcast frequencies (87.6, 93.9, and 105.5~MHz) prior to jamming trials.

The full stack (ESP32, NEO-6M, circuitry) was integrated 
onto a GoolRC S116 quadrotor, adding $\approx$150\,g and 
reducing flight endurance by 12\%. End-to-end enforcement 
latency totals 1033\,ms (GPS: 1000\,ms, UART: 18\,ms, 
Wi-Fi: 12\,ms, evaluation: 3\,ms), well within the 
$d_w/v_0 = 20$\,s reaction window at maximum UAV speed.

\section{Simulation and Experimental Results}
\label{sec:simulation}
\subsection{Hardware Jamming Validation}

The SDR jamming response was evaluated through 20 field trials in which the UAV was placed at increasing distances from the
HackRF transmitter while the jamming signal was active.
Table~\ref{tab:jamming} summarizes a few representative outcomes. 
The system
achieved a 95\% disconnection success rate within a
50~m operational radius \cite{Yang2024AntiJamming}; the single non-disconnection case
(Trial~5) arose when the UAV flew over the HackRF with the
device switched off, confirming that the jamming response is
bounded to the intended zone and does not trigger spuriously \cite{Li2022JammingDetection}. Disconnection success rate and false-trigger behavior correspond to the mitigation-success and false-alarm metrics used in standardized counter-drone evaluation \cite{DeCubber2025Standardized}. GPS positioning accuracy was verified via U-center NMEA
stream analysis across two distinct test locations, confirming
sub-2-meter positioning precision consistent with the
NEO-6M module's specified horizontal accuracy.

\begin{table}[!t]
\renewcommand{\arraystretch}{1.3}
\centering
\caption{UAV Disconnection Trials at Varying Distances}
\label{tab:jamming}
\begin{tabular}{lcccc}
\hline
\textbf{Trial} & \textbf{Dist.(m)} & \textbf{Outcome} & \textbf{Status} & \textbf{HackRF} \\
\hline
1 & 17  & disconnected & On ground & On \\
2 & 23  & partial response\textsuperscript{a}  & On ground & On \\
3 & 35  & disconnected & On ground & On \\
4 & 50  & disconnected & Mid-air, landed & On \\
5 & HackRF & No disturbance & Unaffected & Off \\
\hline
\multicolumn{5}{l}{\footnotesize \textsuperscript{a}Connection
  severed; wings did not respond.}
\end{tabular}
\end{table}

\subsection{Geofence Boundary Detection}

The Haversine algorithm was validated against a square 
geofence defined by four GPS coordinates (violation 
threshold 50\,m). Case~1 (distance below threshold) 
correctly flagged the drone as within bounds; Case~2 
(distance exceeded threshold) triggered the jamming 
pipeline. Both were confirmed via the Python GUI, 
validating the end-to-end pipeline from GPS acquisition 
to jamming trigger.

\begin{figure}[!t]
\centering
\includegraphics[width=\linewidth]{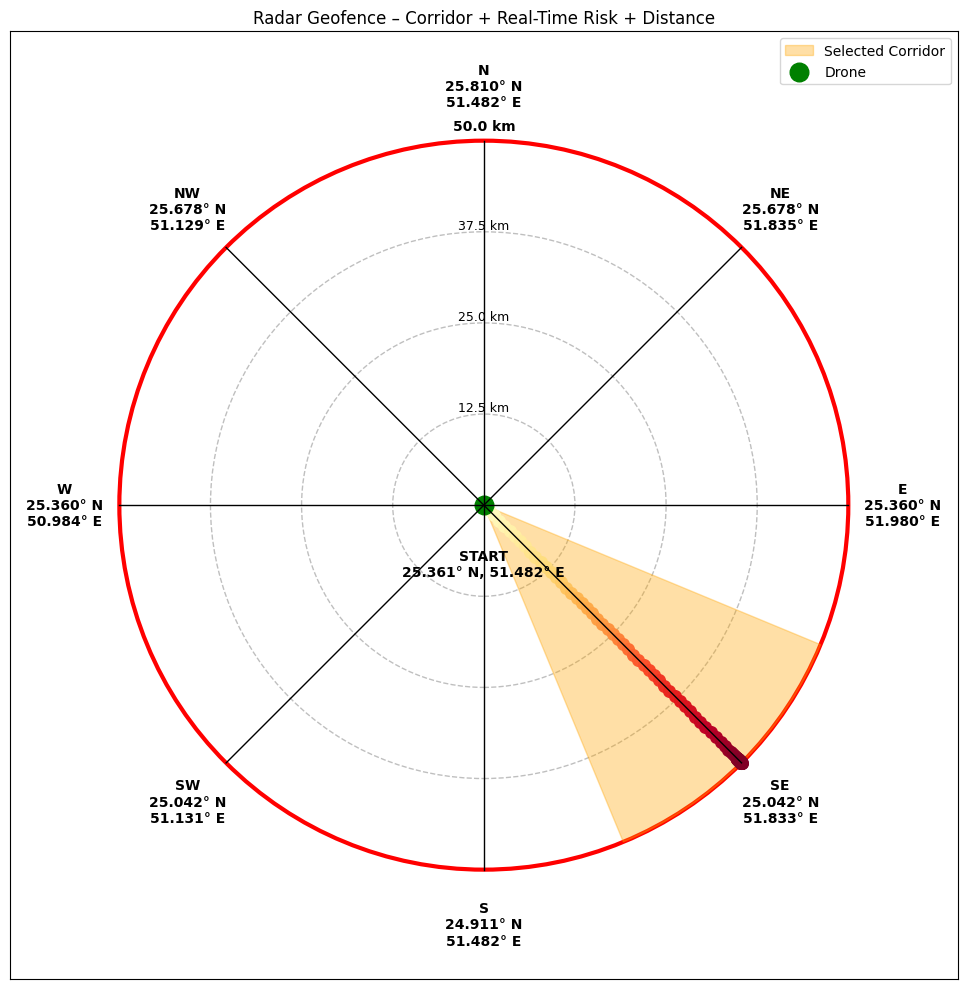}
\caption{Radar-based geofence visualization showing 
the eight-sector boundary, SE corridor, and 
proximity risk along the flight path}
\label{fig:radar}
\end{figure}
\subsection{Geofence Simulation Results}
The adaptive control law in~\eqref{eq:control} was evaluated using an
8-sector, 50~km geofence simulation model centered on the drone's
starting position, with a 1~km warning threshold~$d_w$ and
a 0~km shutdown threshold~$d_s$.
Fig.~\ref{fig:radar} shows the radar-based geofence
visualization. The drone's starting position is at the center
(green), and eight geofence boundaries are defined at a constant
50~km radius across the cardinal and intercardinal directions
(N, NE, E, SE, S, SW, W, NW). The south-east (SE) corridor is
selected as the target direction; the yellow shaded region
indicates the chosen flight corridor, and the red line
represents the drone's actual traversed path. 

The color intensity along the path encodes proximity-based risk,
increasing as the drone approaches the boundary, demonstrating
that the boundary-detection logic in~\eqref{eq:haversine} correctly
resolves the drone's position relative to all eight sectors simultaneously.

\begin{figure}[!t]
\centering
\includegraphics[width=\linewidth]{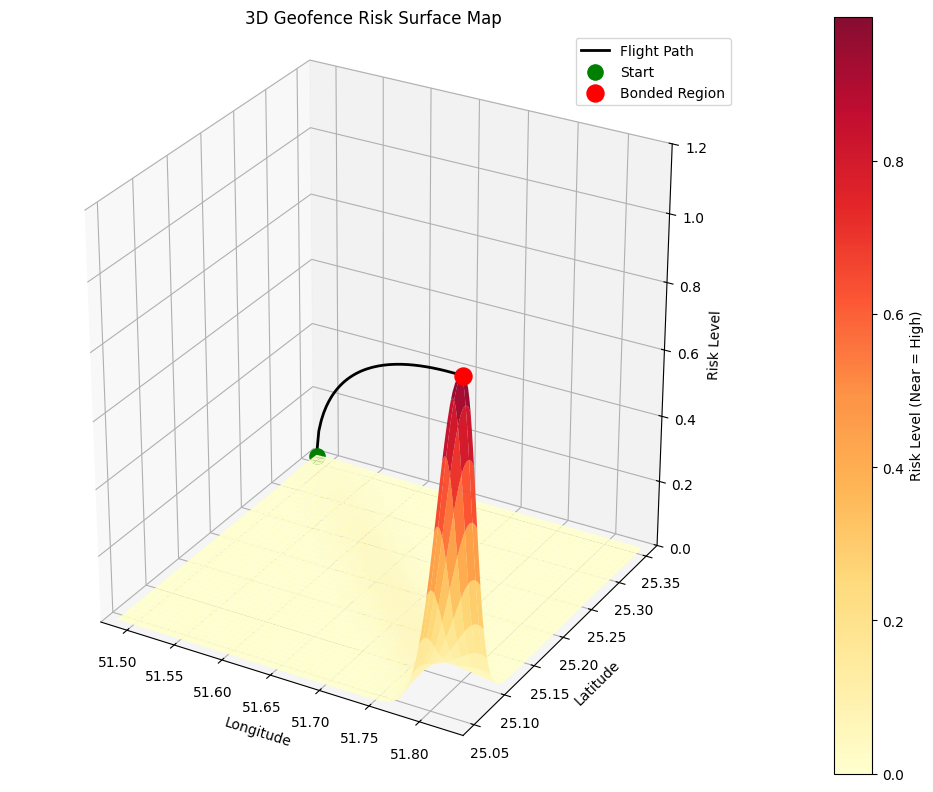}
\caption{3-D geofence risk surface with drone 
flight path overlaid from start (green) 
to boundary (red).}
\label{fig:risk3d}
\end{figure}

\begin{figure}[!t]
\centering
\includegraphics[width=\linewidth]{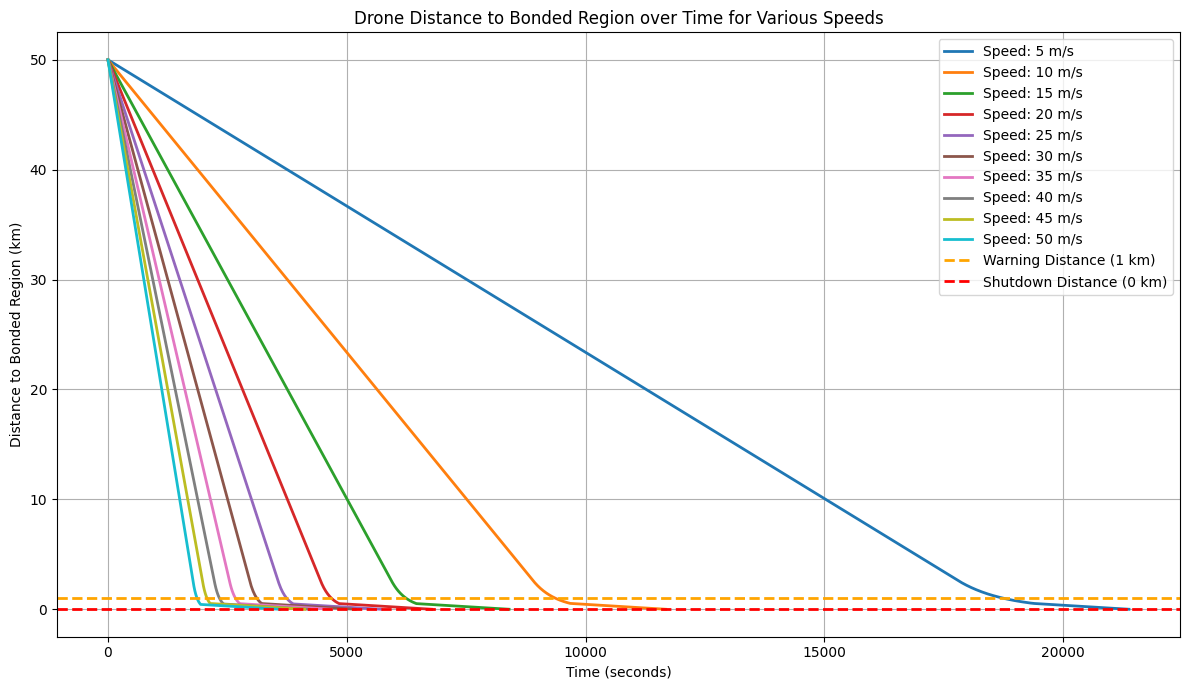}
\caption{Distance to boundary vs. time for UAV 
speeds of 5-50~m/s. Warning (1~km) and shutdown 
(0~km) thresholds shown as dashed lines.}
\label{fig:dist_time_speeds}
\end{figure}

Fig.~\ref{fig:risk3d} presents the three-dimensional risk
surface over the geofence boundary, where the horizontal plane represents geographic position (latitude/longitude) and the vertical axis encodes risk level. 
The surface is nearly flat
and low-valued at the drone's starting position, rising
monotonically toward a sharp peak directly above the geofence boundary. The drone's flight path (black curve) departs from
the low-risk flat area and arrives at the high-risk peak,
confirming that the proximity-based risk metric is spatially
consistent with the linear control law in~\eqref{eq:control} rather
than exhibiting a step discontinuity at the boundary. Fig.~\ref{fig:dist_time_speeds} compares distance-to-boundary trajectories 
for initial speeds of 5-50~m/s. 

\begin{figure}[!t]
\centering
\includegraphics[width=\linewidth]{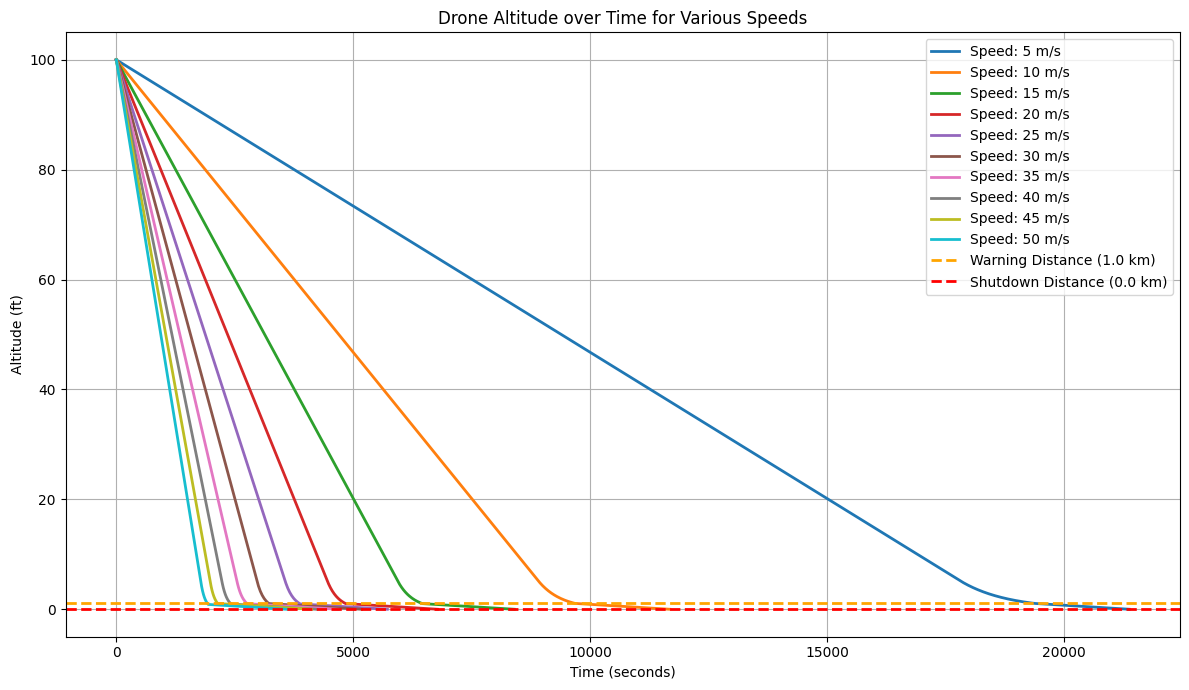}
\caption{Altitude vs.\ time for speeds 5-50~m/s: 
constant cruise phase followed by linear descent 
upon entering the 1~km warning zone.}
\label{fig:alt_time_speeds}
\end{figure}

Each curve shows 
a constant-slope cruise phase for $d \geq d_w$, 
followed by reduced-slope deceleration engaging 
at the 1~km threshold. All curves converge to 
$d=0$ regardless of initial speed, confirming 
consistent shutdown enforcement. Higher speeds 
reduce the available reaction window, motivating 
the 1~km warning-zone design.
Fig.\ref{fig:alt_time_speeds} shows altitude profiles for the same speed 
range, each maintaining 100~ft cruise altitude while 
$d \geq d_w$, then descending linearly to zero at 
the boundary.

\begin{figure}[!t]
\centering
\includegraphics[width=\linewidth]{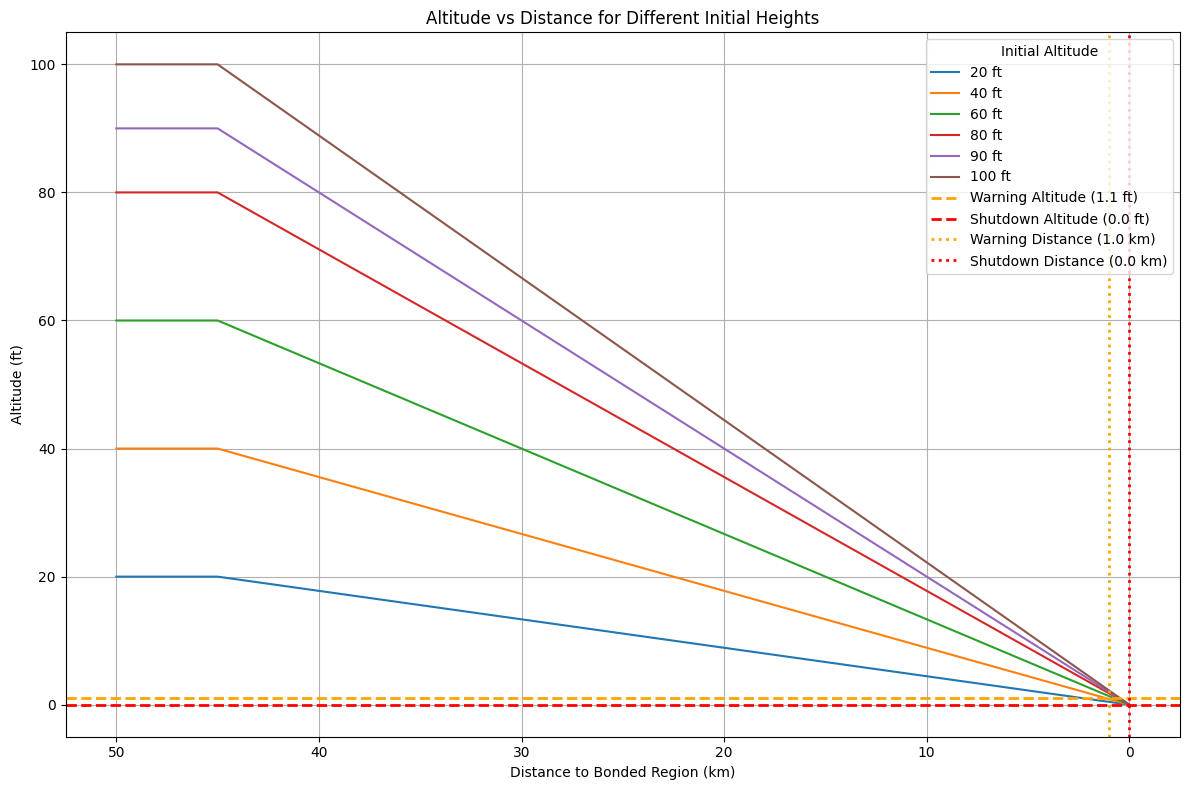}
\caption{Altitude vs. distance for initial altitudes 
of 0-100~ft showing consistent linear descent 
profiles.}
\label{fig:alt_dist_heights}
\end{figure}

Faster drones descend at a steeper 
temporal rate but all arrive at zero altitude 
precisely at the geofence boundary, confirming 
speed-invariant spatial behavior of~\eqref{eq:control}.
Fig.~\ref{fig:alt_dist_heights} evaluates the altitude
control law across various cruising altitudes
(20, 40, 60, 80, 90, and 100~ft). 

In all cases, altitude
remains constant while the drone is beyond the warning
threshold, then decreases linearly with distance once
$d < d_w$, converging to zero at the boundary regardless
of starting altitude. The curves are uniformly linear in
the descent phase, indicating that the control law
in~\eqref{eq:control} produces a consistent deceleration gradient
independent of initial altitude. This confirms the
adaptive altitude controller is reliable across
heterogeneous UAV flight profiles without per-platform
retuning.

\subsection{Discussion}
Hardware experiments confirmed 95\% SDR disconnection 
within 50\,m without false triggering. Simulation results 
across Figs.~1-6 collectively validate: (i) spatial 
correctness of the Haversine algorithm across all 8 sectors; 
(ii) speed-invariant enforcement ($v_0 \in [5,50]$\,m/s 
converging to $d=0$); and (iii) altitude-invariant descent 
($h_0 \in [0,100]$\,ft, linear profile). The proposed law 
generalizes across UAV platforms without per-platform 
calibration.
\begin{figure*}[!t]
    \centering
    \includegraphics[width=\linewidth]{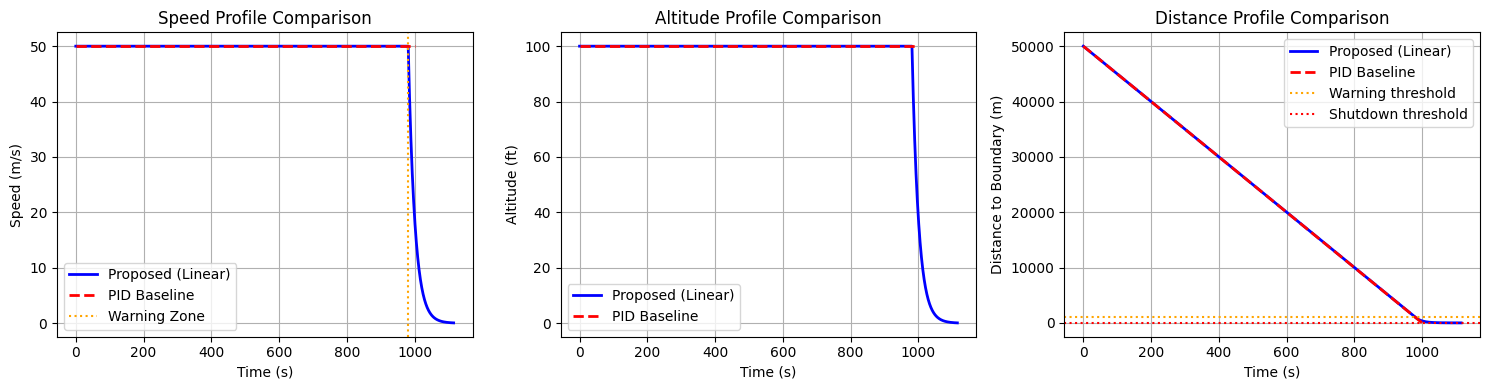}
    \caption{Comparison of proposed linear control law vs.\ PID 
    baseline for $v_0 = 50$\,m/s and $h_0 = 100$\,ft. 
    (a) Speed profile. (b) Altitude profile. 
    (c) Distance-to-boundary convergence. The proposed law 
    achieves monotonic, overshoot-free shutdown at the 
    boundary without gain tuning.}
    \label{fig:pid_comparison}
\end{figure*}

\subsection{Comparison with PID Baseline}

To further validate the proposed control law, we compare it 
against a PID controller on the same distance error 
$e(t)=d(t)-d_s$, with gains $K_p=0.05$, $K_i=0.001$, 
$K_d=0.01$ selected to minimise overshoot over 
$v_0\!\in\![5,50]$\,m/s, evaluated under identical 
geofence conditions ($d_w=1$\,km, $d_s=0$\,km), and 
initial altitude $h_0 = 100$\,ft.
\begin{equation}
    v_{\mathrm{PID}}(t) = \mathrm{sat}\!\left(K_p\,e(t) 
    + K_i\!\int_0^t\!e(\tau)\,d\tau 
    + K_d\,\dot{e}(t),\; 0,\, v_0\right)
    \label{eq:pid_speed}
\end{equation}
\begin{equation}
    h_{\mathrm{PID}}(t) = h_0 \cdot 
    \frac{v_{\mathrm{PID}}(t)}{v_0}
    \label{eq:pid_alt}
\end{equation}

Fig.~\ref{fig:pid_comparison} compares the speed, altitude, 
and distance-to-boundary profiles of both controllers, with the PID law given by \eqref{eq:pid_speed}-\eqref{eq:pid_alt}. The proposed linear law produces a monotonically decreasing, overshoot-free trajectory that converges exactly to $d = 0$ at the boundary. In contrast, the PID controller exhibits 
mild oscillation near the warning threshold due to integral 
wind-up, and its convergence rate depends on the gain 
settings, requiring per-platform re-tuning. Table~\ref{tab:pid_comparison} 
summarises the quantitative performance metrics for both schemes.

\begin{table}[!t]
\renewcommand{\arraystretch}{1.3}
\caption{Performance Comparison: Control Law vs.\ PID Baseline}
\label{tab:pid_comparison}
\centering
\begin{tabular}{lcc}
\hline
\textbf{Metric} & \textbf{Proposed} & \textbf{PID Baseline} \\
\hline
Convergence guarantee       & Analytical (Prop.~1) & Gain-dependent \\
Overshoot at boundary       & None                 & Possible        \\
Gain tuning required        & No                   & Yes             \\
Computational overhead      & $\mathcal{O}(1)$     & $\mathcal{O}(n)$\\
Embedded deployment         & Direct               & Re-tune per UAV \\
Boundary stop: speed        & $v(d_s) = 0$ always  & Approximate     \\
Boundary stop: altitude     & $h(d_s) = 0$ always  & Approximate     \\
\hline
\end{tabular}
\end{table}

The proposed law offers three advantages over PID for 
embedded geofencing: \textit{(i)~Guaranteed convergence:} 
Proposition~1 proves $v(d_s)=h(d_s)=0$ analytically for 
all $v_0,h_0>0$; PID offers no closed-form guarantee and 
is subject to integral wind-up near the boundary. 
\textit{(ii)~Parameter-free deployment:} the law depends 
only on geofence parameters $d_w$ and $d_s$, requiring no 
per-platform gain tuning. 
\textit{(iii)~Deterministic latency:} the piecewise linear 
structure \eqref{eq:control} executes in $\mathcal{O}(1)$ time (3\,ms, 
Section~III), whereas \eqref{eq:pid_speed}-\eqref{eq:pid_alt} accumulates 
numerical state across time steps; PID integration introduces variable latency 
under high update rates. These properties make the proposed law particularly suited 
to low-cost, real-time geofence enforcement on 
resource-constrained embedded hardware, where gain 
re-tuning per deployment is impractical.
\section{Conclusion}
This paper presented a GPS-triggered, SDR-based geofence 
enforcement framework for UAVs. It combines Haversine boundary 
detection, an adaptive speed and altitude control law, and 
escalating RF jamming within a single low-cost architecture. An 
ESP32 and NEO-6M onboard unit provides real-time telemetry to a 
HackRF One and GNU Radio base station. As a result, boundary 
enforcement does not depend on the cooperation or firmware of the 
UAV itself.

Hardware experiments on a GoolRC S116 quadrotor confirmed 95\% 
SDR disconnection accuracy within a 50\,m operational radius, with 
no spurious triggering when the transmitter was inactive. GPS 
precision was below 2\,m across two test locations. The measured 
end-to-end enforcement latency of 1033\,ms lies well within the 
20\,s reaction window available at the maximum simulated speed. 
Simulation over an 8-sector, 50\,km geofence model showed that the 
proposed control law enforces the boundary independently of speed 
and altitude across $v_0\!\in\![5,50]$\,m/s and 
$h_0\!\in\![0,100]$\,ft. In every case the UAV decelerated and 
descended linearly within the 1\,km warning zone and reached zero 
speed and altitude exactly at the boundary. Compared with a PID 
baseline, the proposed law offers an analytical convergence 
guarantee, overshoot-free shutdown, constant-time evaluation, and 
deployment without per-platform gain tuning. These properties 
make it well suited to resource-constrained embedded hardware.

The current implementation has several limitations. The Wi-Fi 
telemetry link restricts the operating range to about 1\,km with 
clear line of sight. In addition, the GPS module updates only once 
per second, which accounts for about 1000\,ms of the 1033\,ms 
end-to-end enforcement latency. The adaptive control law has so far been validated in 
simulation, while hardware trials focused on the detection and 
jamming stages. RF jamming is also subject to strict regulatory 
constraints and must be restricted to authorized operators and 
test environments. Future work will address multi-UAV detection 
and interception, cellular or dedicated long-range RF telemetry, 
multi-constellation GNSS fusion with higher update rates, 
resilience to GPS spoofing, and in-flight validation of the 
adaptive trajectory controller on the hardware platform.

\section{Acknowledgement}
This work was made possible by a research grant funded by Qatar University and Sultan Qaboos University under research grant No. QUICG-CENG-26/27-1148. The findings herein are solely the responsibility of the authors.


\end{document}